\documentclass{article} % For LaTeX2e
\usepackage{icomp2024_conference,times}

\usepackage{amsmath,amsfonts,bm}

\def\eqref#1{equation~\ref{#1}}
\def\1{\bm{1}}

\DeclareMathAlphabet{\mathsfit}{\encodingdefault}{\sfdefault}{m}{sl}
\SetMathAlphabet{\mathsfit}{bold}{\encodingdefault}{\sfdefault}{bx}{n}

\usepackage{hyperref}
\usepackage{url}
\usepackage{booktabs}       % professional-quality tables
\usepackage{amsfonts}       % blackboard math symbols
\usepackage{nicefrac}       % compact symbols for 1/2, etc.
\usepackage{microtype}      % microtypography
\usepackage{xcolor}         % colors
\usepackage{amsmath}
\usepackage[ruled,vlined]{algorithm2e}
\usepackage{algorithm2e}
\usepackage{graphicx}
\usepackage{adjustbox}
\usepackage{subcaption}
\usepackage{amssymb}
\usepackage{multirow}
\usepackage{amsfonts}
\usepackage{cuted,lipsum}
\usepackage[most]{tcolorbox}
\usepackage{subcaption}
\usepackage{tabularx}
\usepackage{graphicx}
\usepackage{hyperref}
\usepackage{multirow}
\usepackage{array}
\usepackage{balance}
\usepackage{makecell}
\usepackage{soul}

\title{SimInterview: Transforming Business Education through
Large Language Model-Based Simulated Multilingual Interview Training System}

\author{Truong Thanh Hung Nguyen$^{1}$$^{*}$$^{\dagger}$ \hspace{1.0em} 
Tran Diem Quynh Nguyen$^{2}$$^{*}$ \hspace{1.0em}  Hoang Loc Cao$^{3}$ 
\AND  
Thi Cam Thanh Tran$^{4}$\hspace{1.0em} 
Thi Cam Mai Truong$^{5}$\hspace{1.0em}
Hung Cao$^{1}$\vspace{1em}\\
$^{1}$Analytics Everywhere Lab, University of New Brunswick, Canada \\
$^{2}$University of Foreign Language Studies, University of Danang, Vietnam\\
$^{3}$Faculty of Information Technology, University of Science, VNU-HCM, Vietnam\\
$^{4}$Faculty of Economics and Accounting, Quy Nhon University, Vietnam\\
$^{5}$Faculty of Natural Sciences, Quy Nhon University, Vietnam\\
}

\icompfinalcopy % Uncomment for camera-ready version, but NOT for submission.
\begin{document}
\def\thefootnote{$^{*}$}\footnotetext{Equal contribution.}
\def\thefootnote{$^{\dagger}$}\footnotetext{Corresponding author: $\texttt{hung.ntt@unb.ca}$}
\maketitle

\begin{abstract} 
Business interview preparation demands both solid theoretical grounding and refined soft skills, yet conventional classroom methods rarely deliver the individualized, culturally aware practice employers currently expect. This paper introduces SimInterview, a large language model (LLM)-based simulated multilingual interview training system designed for business professionals entering the AI-transformed labor market. Our system leverages an LLM agent and synthetic AI technologies to create realistic virtual recruiters capable of conducting personalized, real-time conversational interviews. The framework dynamically adapts interview scenarios using retrieval-augmented generation (RAG) to match individual resumes with specific job requirements across multiple languages. Built on LLMs (OpenAI o3, Llama 4 Maverick, Gemma 3), integrated with Whisper speech recognition, GPT-SoVITS voice synthesis, Ditto diffusion-based talking head generation model, and ChromaDB vector databases, our system significantly improves interview readiness across English and Japanese markets. Experiments with university-level candidates show that the system consistently aligns its assessments with job requirements, faithfully preserves resume content, and earns high satisfaction ratings, with the lightweight Gemma 3 model producing the most engaging conversations. Qualitative findings revealed that the standardized Japanese resume format improved document retrieval while diverse English resumes introduced additional variability, and they highlighted how cultural norms shape follow-up questioning strategies. Finally, we also outlined a contestable AI design that can explain, detect bias, and preserve human-in-the-loop to meet emerging regulatory expectations.
\end{abstract}

\section{Introduction}
\label{sec:intro}
The global labor market is being reshaped by rapid advances in artificial intelligence (AI) and large language models (LLMs) \citep{eloundou_gpts_2024,ghosh_ai-exposure_2025,wilkinson_enhancing_2024}. Business graduates are currently expected to demonstrate not only technical knowledge but also different communication skills that cross disciplinary, linguistic, and cultural boundaries \citep{szkudlarek_communication_2020}. Conventional classroom exercises and one-off mock interviews rarely provide the sustained, personalized practice required to meet these higher expectations. 

Recent LLM breakthroughs offer a pathway to scalable interview coaching, yet most existing tools remain monolingual, text-only, or limited to static question banks \citep{anaza_improving_2023,eysenbach_role_2023,gorer_generating_2023,pasaribu_development_2024}. They seldom integrate the full stack of multimodal AI components, such as speech recognition, voice synthesis, and photorealistic talking heads, needed to approximate the pacing, prosody, and subtle interpersonal cues of live recruitment. Moreover, without retrieval-augmented generation (RAG) grounded in each candidate’s resume and the target job description (JD), systems struggle to produce context-aware dialogue and actionable feedback. Additionally, preparing for interviews in English and Japanese business contexts poses distinct challenges: the former values individual achievement and flexible resume formats, whereas the latter emphasizes long-term organizational fit and the highly standardized resume \citep{sekiguchi2006organizations,moriguchi_japanese_2006,ono_lifetime_2010,aiko2021relationship,pilz_recruitment_2023}. These contrasts can leave otherwise well-qualified candidates under-prepared for culturally specific questioning techniques, follow-up inquiries, and real-time behavioral assessments.

To address these gaps, we propose SimInterview, a multilingual interview-training system that combines a multimodal LLM agent with advanced speech and avatar technologies. Built on LLMs, speech-to-text (STT), text-to-speech (TTS), diffusion-based talking heads, and the vector database, SimInterview delivers real-time, personalized practice sessions that mirror hiring scenarios in both English and Japanese. The framework operates within a modular, privacy-preserving architecture that indexes candidate documents on-premises, retrieves relevant evidence in milliseconds, and then adapts questioning strategy in response to each answer. Our main contributions can be summarized as follows:
\begin{itemize}
    \item \textbf{SimInterview}: LLM-Based Simulated Multilingual Interview Training System that fuses LLM reasoning, low-latency speech processing, and virtual photorealistic avatar rendering to create realistic, conversational interview simulations. Retrieval-augmented personalization that matches resume content with JD requirements, generating targeted questions and culturally sensitive follow-ups in multiple languages.
    \item Evaluation protocols combine automatic alignment metrics with binary user-experience ratings from 20 candidates, showing Gemma 3’s superior conversational quality despite a smaller parameter count than OpenAI o3.
    \item Contestable AI design principles for interview training systems, outlining explainability, auditability, bias mitigation, and human-in-the-loop oversight required under emerging regulations such as the EU AI Act \citep{neuwirth_eu_2022}.
\end{itemize}

By situating simulated interview coaching within an end-to-end, multilingual AI pipeline, SimInterview bridges the persistent gap between academic business education and the dynamic expectations of an AI-driven labor market.
\section{Related Work}
The development of multilingual LLM-based interview training systems sits at the intersection of several rapidly evolving technological domains. This literature review examines recent advances across four critical areas that inform the design and implementation of SimInterview: speech processing technologies, virtual avatar systems, and LLM-based multilingual interview training applications in business education. While significant progress has been made in each domain independently, substantial gaps remain in their integration for multilingual educational applications.

\subsection{Text-to-Speech (TTS) and Speech-to-Text (STT)}
The field of speech synthesis with TTS and SST has undergone a significant transformation with the integration of deep learning (DL) approaches and large-scale language models.
Modern TTS systems have achieved near-human quality in speech generation, with autoregressive models demonstrating the ability to produce speech that is virtually indistinguishable from human speech. Multilingual TTS has emerged as a critical area of research, with models like GPT-SoVITS enabling natural speech synthesis across different languages while preserving appropriate intonation patterns and emotional tone \citep{rvc-boss_rvc-bossgpt-sovits_2025}. The integration of conversational context into TTS systems has been particularly important for applications requiring dynamic speech generation, with models like M2-CTTS demonstrating the ability to synthesize speech with contextually appropriate prosody based on conversation history \citep{xue_m_2023}. The advancement of STT technology has been equally impressive, with OpenAI's Whisper model establishing new benchmarks for multilingual speech recognition \citep{radford_robust_2023}. Its encoder-decoder framework supports real-time segmentation and speaker diarization, critical for dynamic conversational applications like interview simulations.

\subsection{Virtual Avatar Systems}
The development of realistic virtual avatars has been revolutionized by advances in diffusion models and neural rendering techniques. Modern talking head generation systems have achieved remarkable fidelity in creating synchronized facial animations that address the uncanny valley effect often present in earlier avatar-based systems \citep{guo_liveportrait_2024,he_co-speech_2024}.

Recent advances in virtual avatar synthesis have shifted toward diffusion-based architectures that address critical challenges in real-time controllability and temporal coherence. Ditto used a motion-space diffusion framework that explicitly bridges motion generation and photorealistic neural rendering through identity-agnostic representations \citep{li_ditto_2024}. This architecture enables fine-grained control over facial semantics while achieving real-time inference with low first-frame latency, critical for interactive applications like interview simulations. Other innovations include MoDiTalker's motion-disentangled architecture, which separates audio-to-lip synchronization (AToM) from motion-to-video rendering (MToV) to improve lip synchronization accuracy \citep{kim_moditalker_2025}. Meanwhile, IF-MDM introduces implicit face motion modeling to capture fine-grained expressions while balancing motion intensity against visual quality during inference \citep{yang_if-mdm_2024}.

\subsection{Large Language Model (LLM)-Based Interview Training Systems}
The past few years have seen rapid progress in LLMs that drive conversational AI. The evolution of LLMs has been marked by significant increases in model size and capability, with recent models, such as OpenAI’s GPT-4o \citep{openai_gpt-4_2023}, OpenAI’s o3 \citep{openai_introducing_2025}, Meta’s Llama 4 \citep{meta_llama_2025}, and Google’s Gemma 3 \citep{gemma_team_gemma_2025}, achieving state-of-the-art (SOTA) performance across diverse domains. OpenAI's latest models, including GPT-4o for multimodal understanding and o3 for enhanced reasoning, demonstrate SOTA performance in complex tasks like medical diagnostics and algorithmic problem-solving. Meta’s Llama 4 represents the latest advancement in the open-source large language model ecosystem, incorporating Mixture-of-Experts (MoE) architectures that provide superior efficiency-performance trade-offs by activating only specific model components per token. While Google’s Gemma 3 model family represents a significant advancement in lightweight, SOTA open language models, building upon the Gemma 2 architecture with enhanced transformer modifications, including interleaving local-global attention mechanisms and group-query attention.

In business education and career development programs, there is a strong demand for tools that help learners practice job interviews, especially in a global context. Traditional mock interviews with human coaches can be resource-intensive and limited in availability. Also, traditional business education emphasizes functional specialization rather than interdisciplinary problem-solving, with limited integration of practical interview skills in standard curricula \citep{eloundou_gpts_2024,ghosh_ai-exposure_2025,wilkinson_enhancing_2024}. Employers report that graduates lack practical interview skills despite strong academic performance, indicating insufficient preparation for diverse interview formats and cultural contexts. 

The integration of LLMs into interview training has introduced new possibilities for personalized feedback generation, which combines with speech synthesis and virtual avatar technologies to create interactive interview practice experiences. 
\citet{gorer_generating_2023} proposed an automated approach for generating requirements elicitation interview scripts using LLMs through prompt engineering, contributing a graph representation of interactive interview scripts and techniques to generate educational scripts that incorporate typical analyst mistakes to enhance requirements engineering training.
\citet{pasaribu_development_2024} developed an automated LLM-based interview system for information technology (IT) talent recruitment that conducts both behavioral interviews using the STAR (Situation, Task, Action, Result) method and technical interviews, with automated evaluation using fine-tuned Longformer models and RAG from IT domain knowledge. 

However, no comprehensive multilingual AI-powered interview training systems exist that bridge this gap while addressing cross-cultural communication patterns and culturally sensitive feedback mechanisms \citep{anaza_improving_2023,eysenbach_role_2023,gorer_generating_2023,pasaribu_development_2024}. Hence, the identified gaps in cross-cultural multilingual communication modeling and real-time interactive interview systems represent the primary areas where our proposed SimInterview system contributes a novel approach to the field. 

\section{Methodology}
\begin{figure}[]
    \centering
    \includegraphics[width=\linewidth]{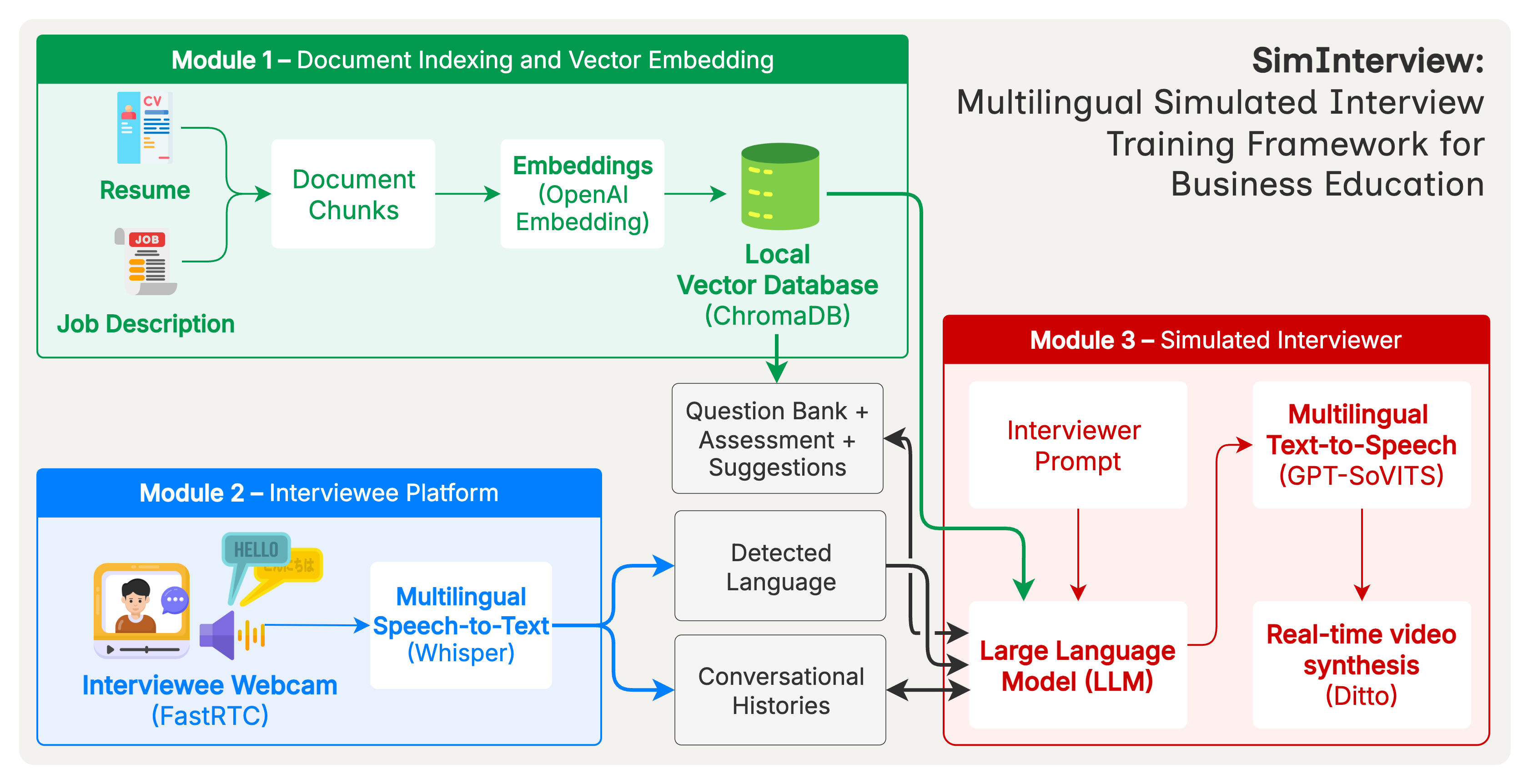}
    \caption{The overview architecture of our proposed SimInterview framework.}
    \label{fig:method}
\end{figure} 

In this section, we describe the overview architecture of SimInterview, which creates a comprehensive interview training system that addresses the gap between theoretical business education and practical interview skills. The framework integrates multiple AI technologies into a cohesive platform that provides personalized, multilingual interview practice tailored to specific job requirements and individual candidate profiles.

Figure~\ref{fig:method} provides the high-level architecture of SimInterview, which transforms unstructured career documents into an interactive, multilingual interview environment. The framework comprises three sequential modules whose outputs propagate through the pipeline, collectively satisfying the design objectives of 
personalization, realism, data confidentiality, low interaction latency, and ensuring scalability while maintaining the flexibility to adapt to various business domains and interview styles as follows:

\begin{figure}[]
    \centering
    \includegraphics[width=\linewidth]{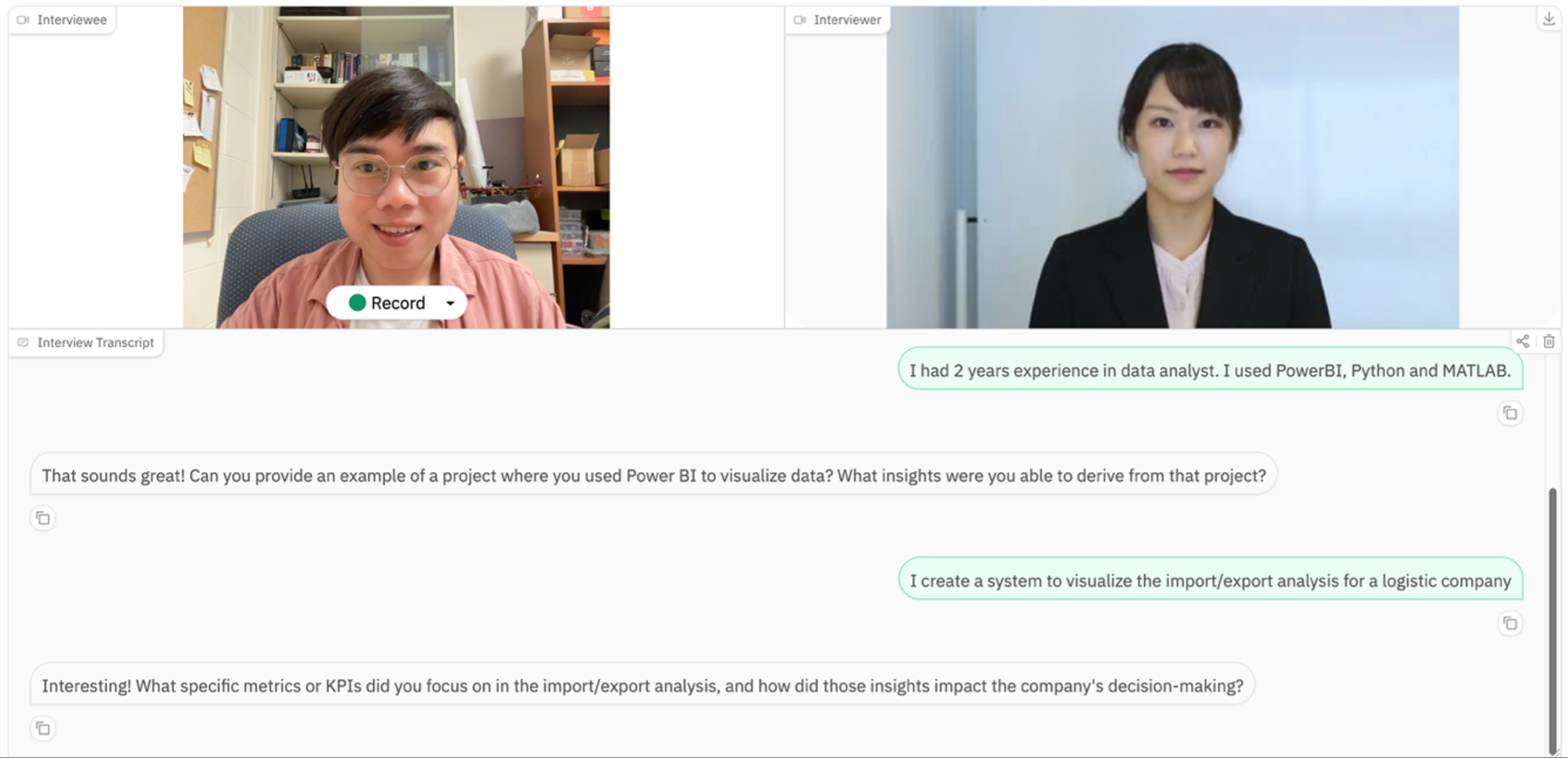}
    \caption{The Real-time Simulated Interviewer interface of our proposed SimInterview framework.}
    \label{fig:demo}
\end{figure}

\begin{itemize}
    \item \textbf{Module 1 – Document Indexing and Vector Embedding:} This module ingests resumes and JDs, removes layout artifacts, normalizes text, and segments it with a sliding window that preserves context across boundaries. Each segment is converted to a unit-norm transformer embedding and stored in an on-premises approximate-nearest-neighbor index, enabling rapid, privacy-preserving retrieval of evidence relevant to any job query.
    \item \textbf{Module 2 – Interviewee Platform:} This module provides the user-facing interface through two primary components: (1) The Resume Assessment and Enhancement component analyzes the alignment between resumes and JDs, generating detailed feedback and improvement suggestions using GPT-4o; (2) Live Interview Interface component manages real-time interactions, including audio capture, speech transcription, and conversation history management. This module creates personalized question banks by identifying relevant experience-requirement matches through vector similarity search, ensuring that interview questions directly relate to both the candidate's background and the target position.
    \item \textbf{Module 3 – Real-time Simulated Interviewer (Figure~\ref{fig:demo}):} This module integrates language generation, speech synthesis, and visual presentation to create a realistic interviewing experience. The module supports multiple LLM backends (OpenAI o3 \citep{openai_introducing_2025}, Llama 4 Maverick \citep{meta_llama_2025}, and Gemma 3 \citep{gemma_team_gemma_2025}) for response generation, each configured with specialized prompts that guide appropriate interviewer behavior. Generated responses undergo TTS conversion using GPT-SoVITS \citep{rvc-boss_rvc-bossgpt-sovits_2025} for natural multilingual speech, while Ditto creates synchronized visual representations through diffusion-based talking head generation. The system maintains conversation context and adapts follow-up questions based on candidate responses, dynamically responding to user input.
\end{itemize}

\section{Implementation}
Our proposed SimInterview framework represents a sophisticated integration of multiple AI technologies to create an immersive interview training environment. Our implementation focuses on three core modules that work synergistically to deliver personalized, multilingual interview experiences while maintaining computational efficiency and user privacy.

\subsection{Module 1 – Document Indexing and Vector Embedding}
The foundation of our personalized interview system lies in its ability to understand and match candidate profiles with job requirements through advanced document processing and semantic embedding techniques. Given an input resume $\mathcal{D}$ and an input JD $\mathcal{J}$, this module transforms unstructured resume and JD data into queryable vector representations that enable intelligent content retrieval during interview sessions.

Our document processing pipeline begins with PyMuPDF for text extraction from various document formats, ensuring compatibility with diverse resume templates and job postings. The extracted text is preprocessed to remove formatting artifacts and normalize character encodings, particularly important for maintaining accuracy across English and Japanese documents. We implement a sliding window approach for text segmentation, employing a chunk size of 512 tokens with a 150-token overlap to preserve contextual continuity across chunk boundaries. This overlapping strategy proves crucial for maintaining semantic coherence when dealing with multi-sentence skill descriptions or experience narratives.

For vector embedding generation, we use OpenAI's \textit{text-embedding-3-small} model, which produces 1536-dimensional dense vectors. The embedding process transforms each text chunk $c_i$ into a vector representation $v_i \in \mathbb{R}^{1536}$ through the function:
\begin{equation}
    v_i=f_\text{embed}(c_i)=\mathrm{normalize}(W \cdot h(c_i)+b)
\end{equation}
where $h\left(c_i\right)$ represents the transformer-based encoding of the text chunk, $W$ is the learned projection matrix, and $b$ is the bias term. The normalization ensures unit vectors for consistent similarity computations.

These embeddings are stored in ChromaDB, a lightweight vector database optimized for similarity search operations. The database employs Hierarchical Navigable Small World (HNSW) graphs for efficient approximate nearest neighbor search. For privacy considerations, the entire vector database operates on a secure local server, eliminating the need for cloud-based storage of sensitive career information. The similarity search mechanism utilizes cosine similarity as the primary distance metric, calculated as:

\begin{equation}
    \mathrm{cosine\_similarity}(v_q, v_d)
    = \frac{v_q \cdot v_d}{\|v_q\| \, \|v_d\|}
    = \frac{\sum_i (v_{q,i} \times v_{d,i})}
            {\sqrt{\sum_i v_{q,i}^2} \times \sqrt{\sum_i v_{d,i}^2}}
\end{equation}
where $v_q$ represents the query vector (derived from JD requirements) and $v_d$ represents document vectors from the candidate's resume. We retrieve the top-k most similar chunks $\mathrm{cosine\_similarity}(v_q,v_d)>\theta$, with a threshold $\theta = 0.75$ to ensure relevance while maintaining diversity in retrieved content.

\subsection{Module 2 – Interviewee Platform}
The interviewee platform serves as the primary interface between candidates and our AI system, designed with careful consideration of user experience and educational effectiveness. This module comprises two interconnected components that guide users through a comprehensive preparation pipeline from resume assessment and enhancement to live interview practice.

\subsubsection{Resume Assessment and Enhancement}
Our resume assessment system employs context-aware analysis powered by OpenAI GPT-4o. The assessment process begins with document standardization, where both resumes and JDs are transformed into structured Markdown representations through prompt templates (Prompt Template~\ref{fig:prompt1} and~\ref{fig:prompt2}). This standardization ensures consistent parsing across varied document formats while preserving hierarchical information such as section headers, bullet points, and emphasis markers.

The structured documents feed into a multi-stage analysis pipeline. First, the system performs section-wise evaluation, examining each resume component (e.g., professional summary, experience, skills, education) against corresponding job requirements. The matching algorithm considers not only explicit keyword overlap but also semantic similarity between stated qualifications and desired competencies. For instance, a resume mentioning ``led cross-functional teams'' would match strongly with a JD requiring ``collaborative leadership experience'' despite the absence of exact phrase matches. Finally, the resume assessment and enhancement engine generates three outputs: competency alignment scores, section-specific feedback, and actionable enhancement recommendations via Prompt Template~\ref{fig:prompt3}.

\subsubsection{Live Interview Interface}
The interview session module represents the technical core of SimInterview, combining multiple AI services to create seamless conversational experiences. Upon session initialization, the system constructs a personalized question bank by analyzing the intersection between candidate qualifications and job requirements. 

This analysis employs our vector similarity search to identify the most relevant experience-requirement pairs, from which contextually appropriate interview questions are generated. Question generation follows a structured approach encoded in JSON format via Prompt Template~\ref{fig:prompt4}, with each question object containing the query text, expected competency areas, and difficulty level. The system maintains balance across behavioral, technical, and situational question types, adapting the distribution based on the specific role requirements. For technical positions, the ratio might favor problem-solving scenarios, while managerial roles emphasize leadership and strategic thinking questions.

Real-time interaction relies on WebRTC technology through FastRTC implementation, enabling low-latency audio-visual communication essential for natural conversation flow. The audio stream from the candidate's microphone is continuously processed through OpenAI's Whisper model \citep{radford_robust_2023}, configured with language detection enabled to support our multilingual capabilities. The transcription pipeline maintains a rolling buffer of conversation history, structured as alternating interviewer-candidate exchanges. This conversational context proves crucial for generating follow-up questions and maintaining topical coherence throughout the session.

\subsection{Module 3 – Real-time Simulated Interviewer}
The simulated interviewer module combines all the above technical components to create a real-time simulated interviewer. This module coordinates language understanding, response generation, and multimodal presentation to provide realistic interview practice that mirrors actual business interview scenarios.

Our experimental framework evaluates three SOTA LLMs as the conversational backbone: OpenAI o3 \citep{openai_introducing_2025}, Llama 4 Maverick \citep{meta_llama_2025}, and Gemma 3 \citep{gemma_team_gemma_2025}. Each model underwent extensive prompt engineering to optimize interview behavior, balancing professional demeanor with appropriate follow-up questioning. Prompt Template \ref{fig:prompt5} incorporates several key elements: role definition, establishing the interviewer persona, contextual grounding using retrieved resume-JD matches, conversation history for maintaining coherence, and behavioral guidelines ensuring appropriate difficulty progression.

Response generation follows a multi-step process. First, the LLM analyzes the candidate's most recent response in conjunction with the conversation history and remaining question bank. It evaluates response completeness, identifies areas requiring clarification, and determines whether to dive deeper or transition to new topics. The decision logic incorporates factors such as response length, keyword coverage, and demonstrated competency levels. For incomplete or vague answers, the system generates targeted follow-up questions, while comprehensive responses trigger progression to new competency areas.

The TTS conversion employs GPT-SoVITS~\citep{rvc-boss_rvc-bossgpt-sovits_2025}, a multilingual synthesis model specifically optimized for English and Japanese. This model generates speech with natural intonation patterns crucial for conveying appropriate emotional tone during interviews. The synthesis process considers punctuation, sentence structure, and contextual emphasis to produce speech that sounds engaged and responsive rather than mechanically generated. 

Visual presentation adds another layer of realism through Ditto, a diffusion-based talking head generator \citep{li_ditto_2024}. This component creates synchronized lip movements and facial expressions corresponding to the generated speech, addressing the uncanny valley effect often present in avatar-based systems. The diffusion model operates at 24 frames per second (FPS), maintaining temporal consistency while generating subtle facial movements that enhance perceived naturalness. The visual generation pipeline processes audio features extracted from the synthesized speech, predicting corresponding facial muscle activations that drive the avatar’s expressions.

Performance optimization remains a critical consideration throughout the system architecture. We implement response caching for frequently asked questions, reducing LLM inference. The complete SimInterview framework achieves end-to-end low latency from candidate speech completion to interviewer response initiation, maintaining the natural rhythm of human conversation. Together, the three modules create an intelligent interview training system where document understanding drives personalization, real-time interaction enables practical skill development, and multimodal AI delivery ensures engaging, realistic practice sessions. This integrated approach transforms static resume preparation into dynamic conversational experiences, helping candidates build confidence and competence for success in the evolving business landscape.

\section{Experiments and Results}
\subsection{Experimental Setup}
To rigorously evaluate the effectiveness of our SimInterview framework, we conducted comprehensive experiments using three LLMs that represent different scales and architectural approaches: OpenAI's o3 (representing large-scale models), Llama 4 Maverick (17B parameters), and Gemma 3 (27B parameters). This diverse model selection enables us to assess framework performance across varying computational capacities and model architectures.

\begin{figure}[ht!]
    \centering
    \includegraphics[width=\linewidth]{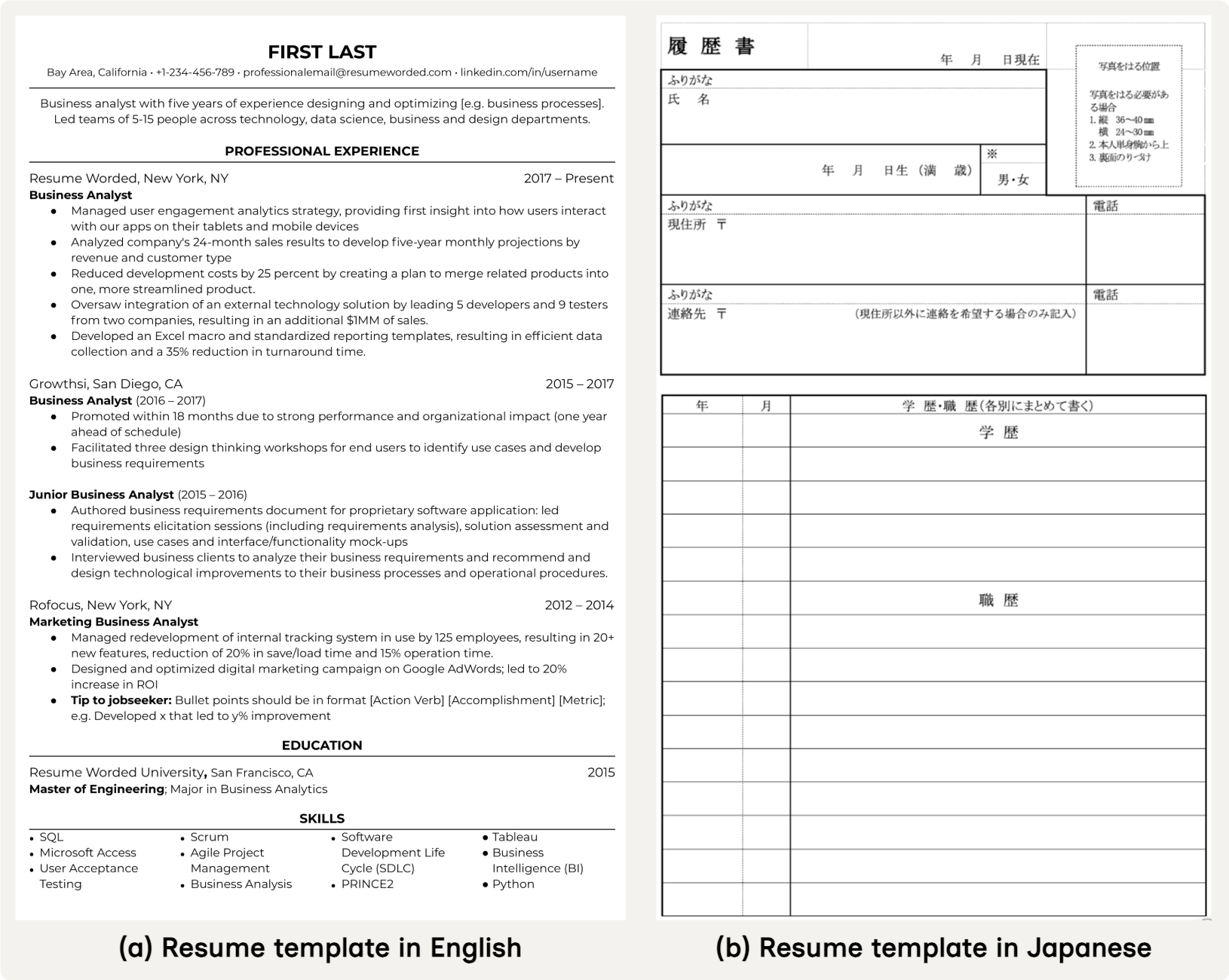}
    \caption{Resume templates in (a) English and (b) Japanese for the evaluation protocols.}
    \label{fig:template}
\end{figure}

Our evaluation focuses on English and Japanese as primary languages, chosen specifically for their fundamental linguistic and cultural differences that present unique challenges for interview simulation systems. These languages differ significantly in their writing systems, i.e., English employs an alphabetic script while Japanese uses a combination of syllabic scripts (\textit{hiragana} and \textit{katakana}) and logographic characters (\textit{kanji}). Beyond linguistic differences, these languages represent distinct interview cultures and professional norms. Most notably, resume and JD formats vary considerably between these contexts, as shown in Figure~\ref{fig:template}. English-language resumes in our dataset show diverse formatting styles, reflecting the flexibility common in Western professional contexts, whereas Japanese resumes predominantly follow the standardized \textit{rirekisho} format, i.e., a highly structured, culturally-specific template that reflects Japan's more formalized approach to professional documentation. All experiments were conducted on an NVIDIA A100 Tensor Core GPU to ensure consistent computational conditions and reproducible results across all model evaluations.

\subsection{Metrics}
We evaluate the framework with two protocols: (1) the resume assessment and enhancement module, and (2) the interview session module by evaluating the user experience while using the SimInterview framework.

\subsubsection{Resume Assessment and Enhancement}
We adopted two metrics to evaluate the resume assessment and enhancement module: assessment alignment and content preservation metrics.
\paragraph{Assessment Alignment (AA)} This metric measures the alignment between the generated resume’s assessment and the JD. The higher score means the assessment is well-tailored to the specific JD the user is applying for. This metric measures in both the token space and the latent space. In the token space, we formulate the AA as the overlap coefficient between the unique words in the generated resume’s assessment and the JD as follows:
\begin{equation}
    \mathrm{AA}_\text{token}=\frac{|\mathcal{W}(\mathcal{A})\cap\mathcal{W}(\mathcal{J})|}{\text{min}\big(|\mathcal{W}(\mathcal{A})|,|\mathcal{W}(\mathcal{J})|\big)}
\end{equation}
where $\mathcal{A}$ is the generated resume’s assessment, $\mathcal{J}$ is the JD, and $\mathcal{W}(\cdot)$ refers to the set of unique words. Similarly, in the latent space, we consider the cosine similarity between the generated resume’s assessment and the JD. Both texts are embedded with a TF-IDF (term frequency–inverse document frequency) vectorizer with stop-words removed, and the cosine similarity between the two vectors is computed as follows:
\begin{equation}
    \mathrm{AA}_\text{latent}=\mathrm{cosine\_similarity}(\mathcal{E}(\mathcal{A}),\mathcal{E}(\mathcal{J}))
\end{equation}
where $\text{cosine\_similarity}(\cdot,\cdot)$ refers to the cosine similarity, $\mathcal{E}(\cdot)$ refers to the encoding used to embed the text.

\paragraph{Content Preservation (CP)} Since hallucination is a known issue even in SOTA LLMs, we would want to verify how well our proposed SimInterview pipeline is able to preserve content between the user-provided resume and the generated one. Ideally, we would want high values for this, since low values of content preservation might imply hallucination by the LLM. Similar to the AA metric, we measure in both the token space and the latent space. In the token space, we formulate this as the overlap coefficient between the generated resume’s assessment and the candidate’s resume:
\begin{equation}
\mathrm{CP}_{\text{token}} =
\frac{\lvert \mathcal{W}(\mathcal{A}) \cap \mathcal{W}(\mathcal{D}) \rvert}
{\min\big( \lvert \mathcal{W}(\mathcal{A}) \rvert, \lvert \mathcal{W}(\mathcal{D}) \rvert \big)}
\end{equation}
where $\mathcal{D}$ denotes the original resume input by the candidate. Similarly, the CP metric in the latent space is formulated as:
\begin{equation}
    \mathrm{CP}_\text{latent}=\mathrm{cosine\_similarity}(\mathcal{E}(\mathcal{A}),\mathcal{E}(\mathcal{D}))
\end{equation}

The combination of these two metrics provides an idea of how good the generated resume’s assessment is. The best scenario is when both metrics are high, which means the LLM assessment is faithful and accurate. In this evaluation pipeline, we evaluate 10 pairs of resumes and JDs in each language.

\subsubsection{Interview Session User Experience}
\begin{figure}
    \centering
    \includegraphics[width=\linewidth]{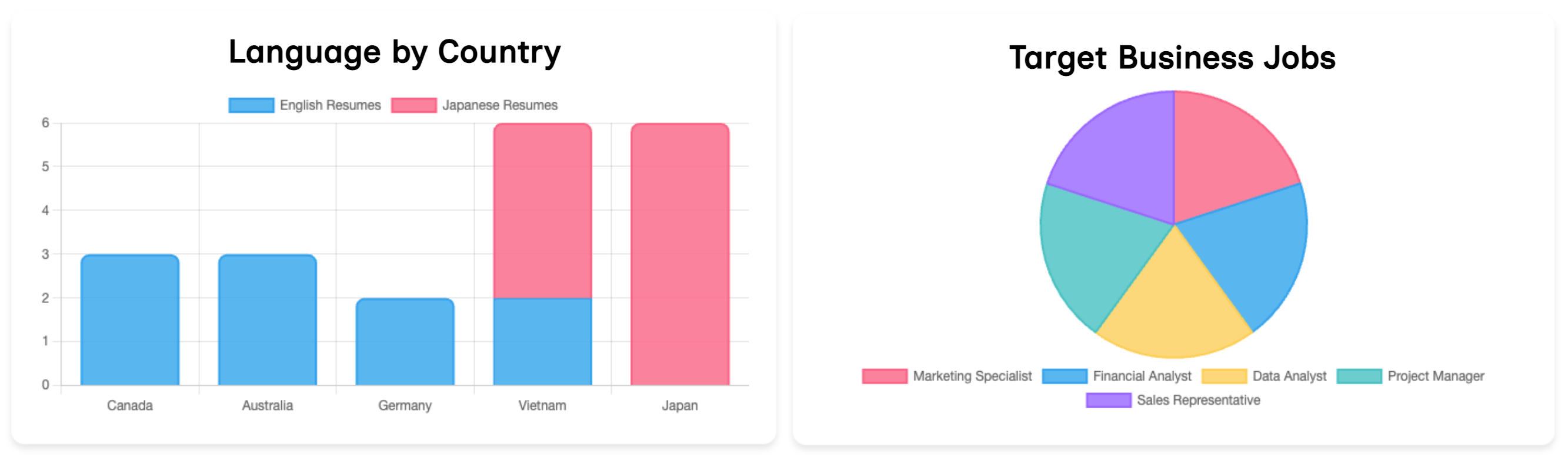}
    \caption{The demographics of candidates. Distribution of candidates by (a) language and country of origin, and (b) target business job categories.}
    \label{fig:demographics}
\end{figure}

Our evaluation pipeline employed a comprehensive user experience assessment protocol designed to measure the effectiveness and user satisfaction of the SimInterview framework across diverse demographic and linguistic populations. The study utilized a mixed-methods approach combining real-time user feedback collection with structured interview simulations. We recruited 20 candidates aged 18-27 years from multiple countries, ensuring representation across different linguistic backgrounds and professional aspirations. Participants were selected based on their native or advanced proficiency in either English or Japanese, with expertise levels validated through a preliminary language assessment. The candidate pool encompassed individuals pursuing various business career paths, including but not limited to marketing, finance, consulting, technology, and international business. Detailed demographic distributions, including country of origin, language proficiency levels, educational backgrounds, and target industry preferences, are presented in Figure~\ref{fig:demographics}. Each participant engaged in a standardized interview simulation lasting approximately 15 minutes, conducted entirely within the SimInterview system using their preferred language. 

Throughout the interview process, we implemented a binary feedback collection system where participants provided immediate responses after each dialogue exchange. The binary scale ($\text{Like}=+1$, $\text{Dislike}=-1$) was chosen to minimize cognitive load during the interview while ensuring consistent, quantifiable data across diverse cultural contexts where rating scale interpretations may vary. This real-time feedback capture allowed us to assess user experience at granular interaction levels rather than relying solely on post-interview retrospective evaluations. The primary evaluation metric was calculated as the arithmetic mean of all binary feedback values per participant, providing an overall satisfaction score ranging from $-1$ (completely negative experience) to $+1$ (completely positive experience). This aggregated score serves as our key indicator for system effectiveness and user satisfaction, enabling comparative analysis across demographic groups, language preferences, and target job categories.

\subsection{Evaluation Results}
Table~\ref{tab:results} and Figure 5 present the performance comparison of three SOTA LLMs (i.e., OpenAI o3 \citep{openai_introducing_2025}, Llama 4 Maverick \citep{meta_llama_2025}, and Gemma 3 \citep{gemma_team_gemma_2025}) across both English and Japanese interview simulations. The evaluation includes assessment alignment ($\mathrm{AA}$), content preservation ($\mathrm{CP}$), and user experience metrics, with all metrics scored on a scale where higher values indicate better performance.

\begin{table}[t]
\centering
\caption{LLM performance comparison for English/Japanese interviews (higher values = better performance; best results in \textbf{bold}).}
\label{tab:results}
\begin{tabular}{llccccc}
\toprule
\multirow{2}{*}{Language} & \multirow{2}{*}{Model} & \multicolumn{2}{c}{$\mathrm{AA}$} & \multicolumn{2}{c}{$\mathrm{CP}$} & \multirow{2}{*}{User Experience} \\
\cmidrule(lr){3-4} \cmidrule(lr){5-6}
 & & (token) & (latent) & (token) & (latent) &  \\
\midrule
\multirow{3}{*}{English}
 & OpenAI o3 & \textbf{0.605} & \textbf{0.549} & 0.712 & 0.664 & 0.755 \\
 & Llama 4   & 0.578 & 0.513 & 0.689 & 0.642 & 0.703 \\
 & Gemma 3   & 0.598 & 0.532 & \textbf{0.722} & \textbf{0.682} & \textbf{0.812} \\
\midrule
\multirow{3}{*}{Japanese}
 & OpenAI o3 & \textbf{0.599} & \textbf{0.534} & 0.742 & 0.698 & 0.651 \\
 & Llama 4   & 0.552 & 0.491 & 0.645 & 0.602 & 0.621 \\
 & Gemma 3   & 0.579 & 0.512 & 0.732 & 0.687 & \textbf{0.787} \\
\bottomrule
\end{tabular}
\label{tab:results}
\end{table}

\begin{figure}[t]
    \centering
    \includegraphics[width=\linewidth]{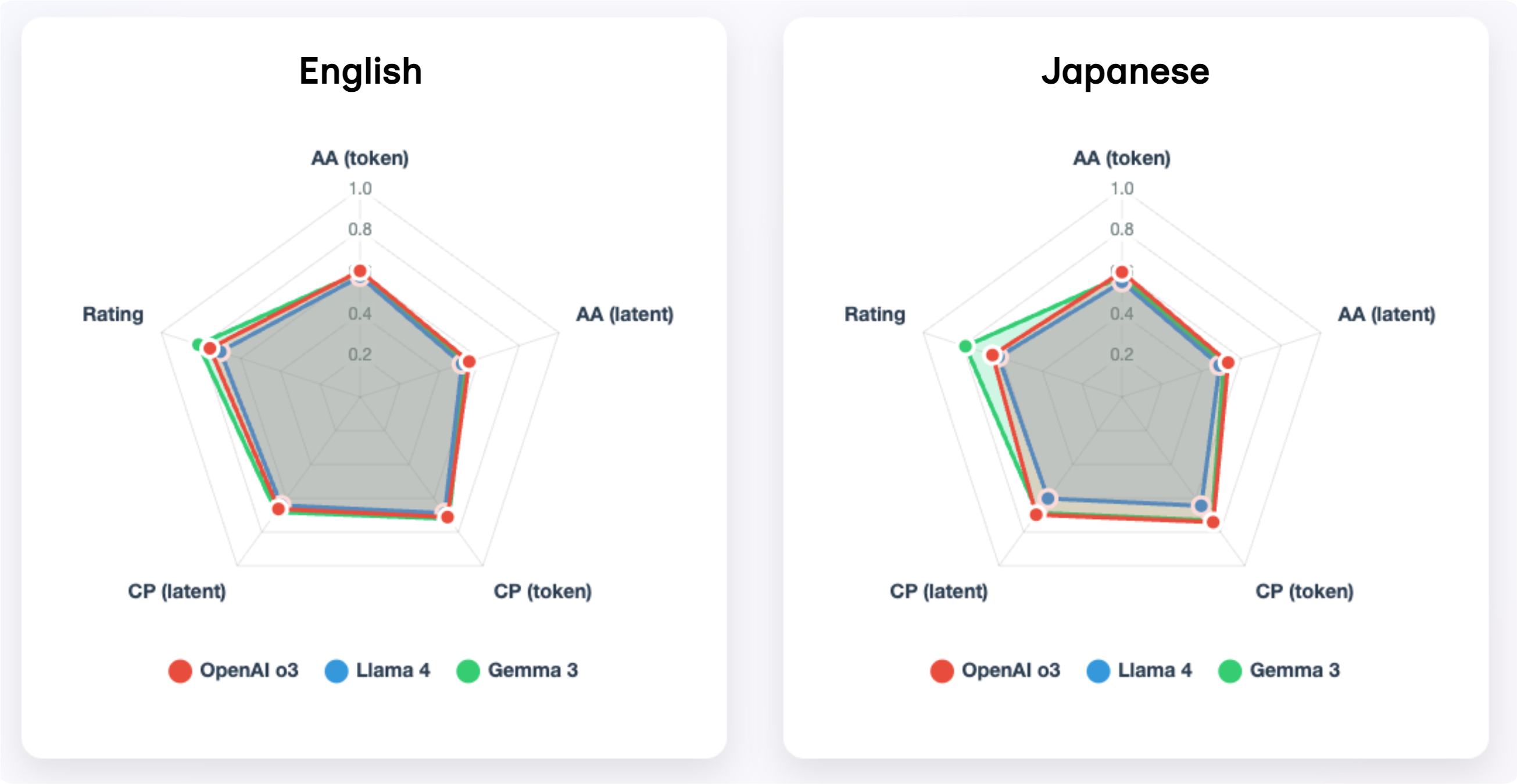}
    \caption{The spider plots of LLM performance comparison for English/Japanese interviews.}
    \label{fig:placeholder}
\end{figure}

Across both languages, Gemma 3 demonstrates superior user experience scores, achieving the highest ratings in both English (0.812) and Japanese (0.787) interviews. This suggests that Gemma 3 generates more engaging and satisfactory interview interactions from the candidates' perspective. OpenAI o3 shows competitive performance in assessment alignment metrics, particularly excelling in token-level assessment alignment for English interviews (0.605).

The results reveal an interesting relationship between model size and performance effectiveness. Llama 4, the smallest model with 17B parameters, consistently underperforms across all metrics in both languages, suggesting that parameter scale plays a crucial role in interview simulation quality. However, the performance gap between Gemma 3 (27B) and OpenAI o3 (the largest model) demonstrates that raw parameter count does not guarantee superior performance across all dimensions.
\section{Discussion}
In this section, we discuss the impact of cultural differences on LLM-based interview training systems and examine the importance of implementing contestable AI principles in interview training systems.

\subsection{Impact of Recruitment Cultures on LLM-Based Interview Training Systems}
Extensive peer-reviewed research documents systematic differences between Japanese and Western countries using English as the primary language for recruitment approaches. Japanese corporate culture is fundamentally shaped by traditional values that emphasize collective harmony and family-like organizational structures. Academic research reveals that Japanese national culture has been integrated with corporate culture, with the concept of family-based thinking, people-oriented ideas, and loyalty spirit being embodied in corporate practices \citep{piao_case-based_2019}. Japanese companies prioritize long-term cultural integration and family stability assessment, as Japanese companies view employees as complete persons whose family stability affects organizational stability, while Western firms emphasize immediate competency demonstration and individual achievement \citep{pilz_recruitment_2023}. Furthermore, the lifetime employment system fundamentally shapes Japanese interview approaches by creating long-term relationship expectations rather than transactional skill exchanges \citep{moriguchi_japanese_2006,ono_lifetime_2010}. Academic studies show Japanese companies hire for potential and cultural integration over specific skills, reflecting the understanding that technical competencies can be developed through company-specific training programs while cultural fit cannot be easily taught, emphasizing homogeneous organizational culture maintenance and long-term development investment \citep{koyama_mutual_2022}. Hence, although LLMs have the capability to personalize the question banks for personal and cultural aspects, the high diversity in questions that can be raised in the Japanese interviews makes it harder for LLMs to simulate the real-world scenario in Japanese culture, which reflects the lower user experience rating for LLMs in Japanese interviews than English ones.

On the other hand, as shown in Figure~\ref{fig:template}, the standardized \textit{rirekisho} format of Japanese resumes creates significant advantages for LLM performance in RAG systems designed for interview training, contrasting sharply with the challenges posed by diverse English resume formats. As shown by the CP metric in Table~\ref{tab:results}, LLMs consistently performed better when processing Japanese resumes than English ones, suggesting that the generated assessments more accurately reflected the content of the original input resumes in the Japanese context. Our results showed that consistent formatting improves parsing accuracy for LLM-based systems.

\subsection{Towards Contestable AI Interview Training Systems}
The development of AI-based interview training systems like SimInterview demands careful consideration of contestable AI principles by emphasizing systems that are open and responsive to dispute throughout their entire lifecycle \citep{ploug2020four,alfrink_contestable_2023,nguyen_heart2mind_2025}. For interview training systems, this paradigm shift is particularly critical given the human rights, high-stakes nature of professional development, and the potential for algorithmic bias to perpetuate discriminatory hiring practices \citep{hunkenschroer_is_2023}.

The implementation of contestable AI principles in interview training systems presents both significant opportunities and complex challenges. Recent advances in argumentative LLMs demonstrate promising approaches for creating contestable LLM-based systems through formal argumentative reasoning frameworks that enable argument strength contestability and transparent explanations derived from deterministic procedures \citep{clement2023coping,freedman_argumentative_2024,nguyen_heart2mind_2025,nguyen2025congait}. However, the application of these principles to multilingual interview training contexts introduces additional complexity around cultural sensitivity and cross-cultural communication patterns. Research in educational AI systems reveals that while explainable feedback mechanisms can enhance learning outcomes and build trust \citep{thai_educational_2024}, they must be carefully designed to preserve user agency and avoid AI dependency that undermines self-regulated learning strategies \citep{zhang_conversational_2025}. For SimInterview, this suggests the need for adaptive explanation mechanisms that can accommodate diverse cultural contexts while maintaining learner autonomy in the training process.

The regulatory landscape further underscores the importance of contestable design in interview training systems. The EU AI Act classifies AI systems used in education and vocational training as high-risk, mandating transparency obligations, human oversight measures, and fundamental rights impact assessments \citep{neuwirth_eu_2022}. For LLM-based interview training systems, these requirements translate to specific technical implementations, including comprehensive audit trails, human-in-the-loop oversight mechanisms, and bias detection systems that can identify and mitigate discriminatory outcomes across different demographic groups and cultural contexts. For example, technical implementations should include interactive explanation mechanisms that allow trainees to understand and challenge AI-generated feedback \citep{nguyen_langxai_2024}, modular system architectures that can be modified based on successful contestation, and comprehensive logging systems that enable external assessment \citep{nguyen2025xedgeai}.

Organizationally, this requires establishing clear responsibility chains for AI decisions, creating escalation pathways for unresolved challenges, and implementing continuous monitoring systems to assess both system performance and contestability effectiveness. As the field evolves toward more sophisticated human-AI collaboration frameworks, interview training systems like SimInterview have the opportunity to pioneer contestable AI approaches that not only improve training outcomes but also contribute to more fair and transparent AI deployment in professional development contexts.

\section{Conclusion}
In summary, SimInterview demonstrates that a retrieval-augmented, multimodal LLM pipeline can deliver realistic, culturally adapted interview practice at scale, measurably improving candidate confidence and performance in both English and Japanese contexts. Empirical results confirm strong assessment alignment, high content preservation, and favorable user-experience scores, especially for the lightweight Gemma 3 model. Together, these findings position SimInterview as an effective, extensible bridge from classroom to boardroom and as a foundation for future research on multilingual, responsible, and contestable AI interview training systems.

\bibliography{ref}
\bibliographystyle{icomp2024_conference}

\clearpage
\appendix
\section{Appendix}

\renewcommand{\figurename}{Prompt Template} % Change "Figure" to "Prompt Template"
\setcounter{figure}{0} % Restart numbering at 0 so first one is 1

\begin{figure}[th!]
    \centering
    \includegraphics[width=\linewidth]{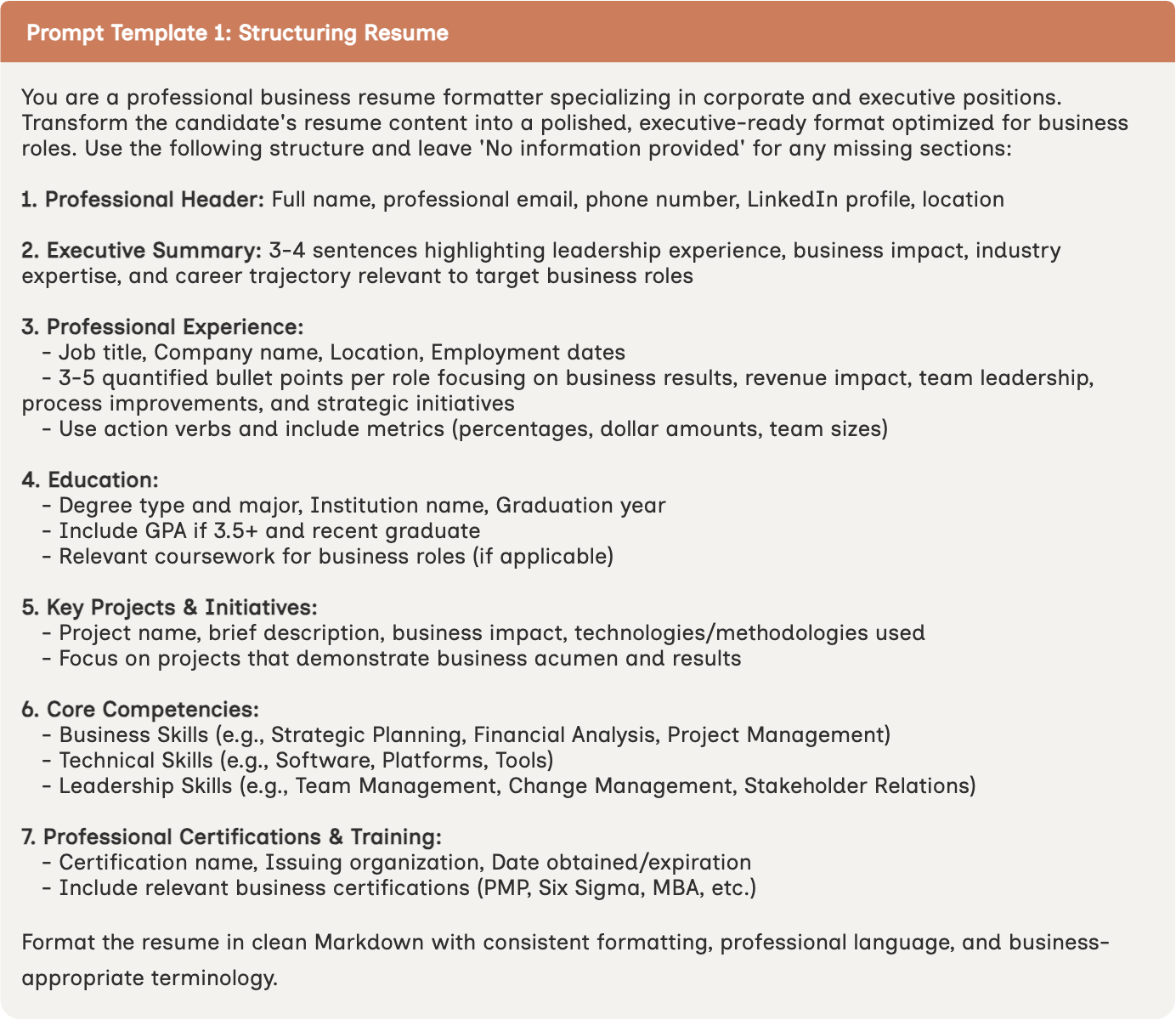}
    \caption{Structuring the resume in the Markdown format.}
    \label{fig:prompt1}
\end{figure}
\clearpage
\begin{figure}[th!]
    \centering
    \includegraphics[width=\linewidth]{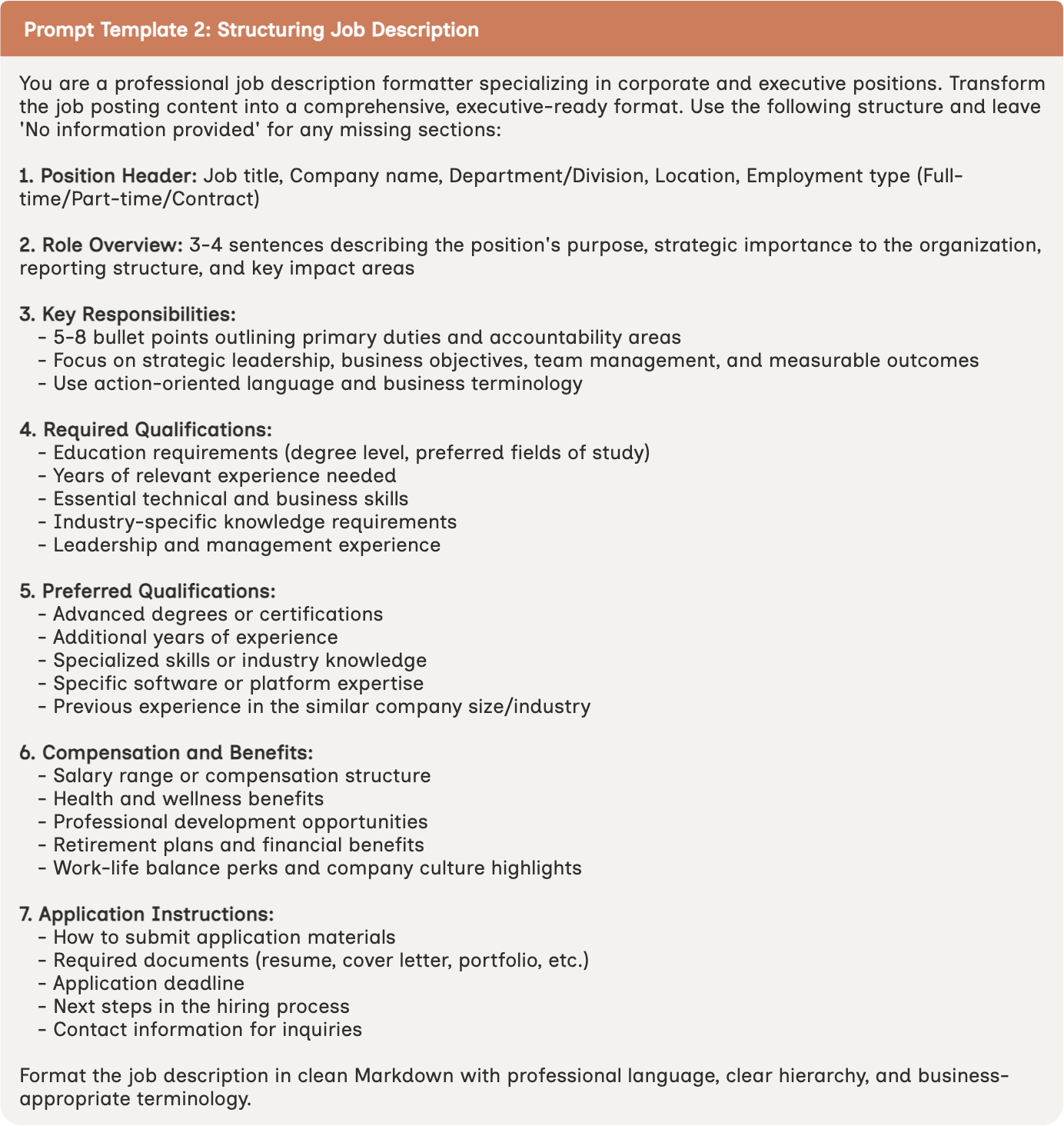}
    \caption{Structuring the JD in the Markdown format.}
    \label{fig:prompt2}
\end{figure}
\clearpage
\begin{figure}[th!]
    \centering
    \includegraphics[width=\linewidth]{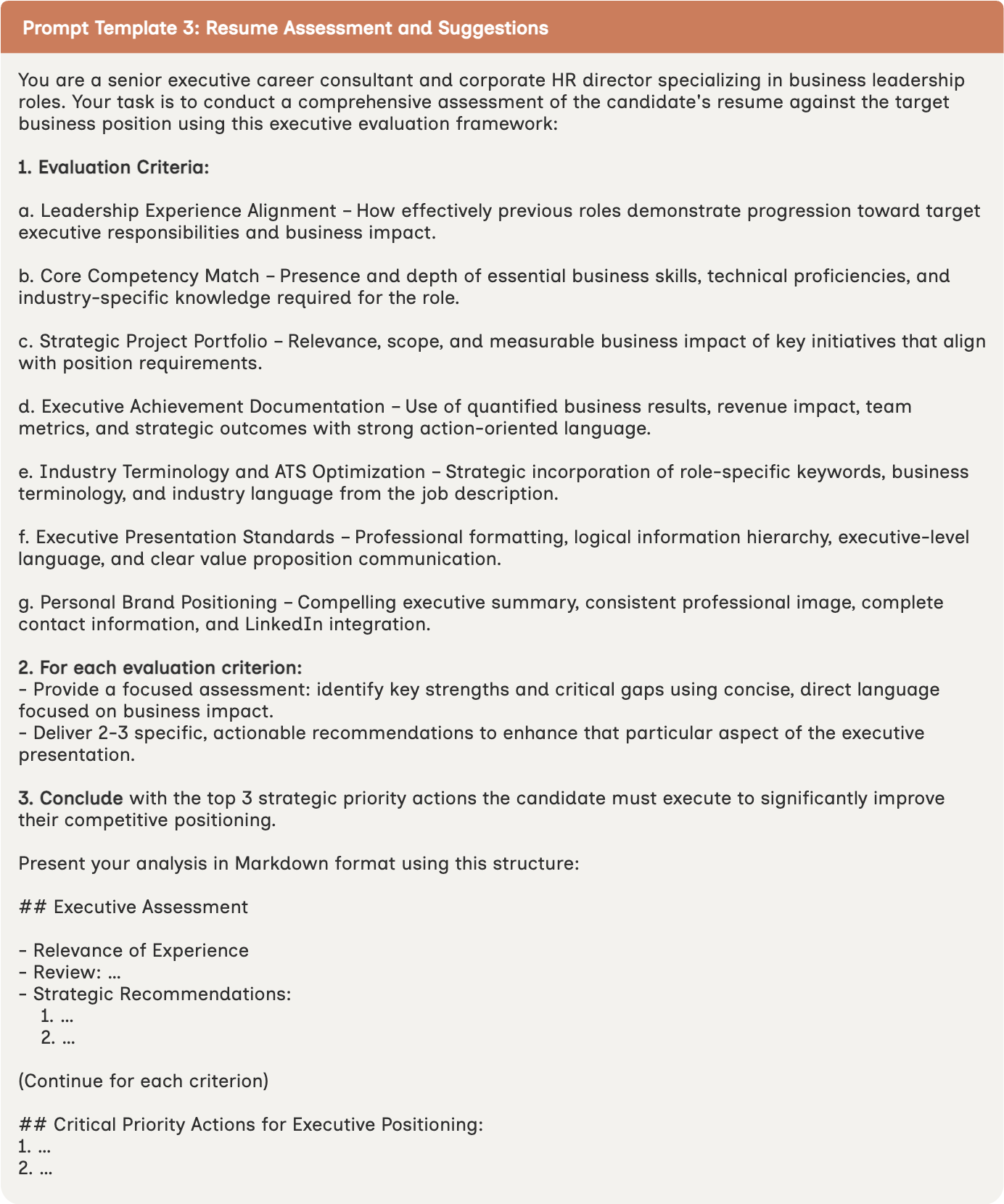}
    \caption{Resume assessment, review, and enhancement suggestions.}
    \label{fig:prompt3}
\end{figure}
\clearpage
\begin{figure}[th!]
    \centering
    \includegraphics[width=\linewidth]{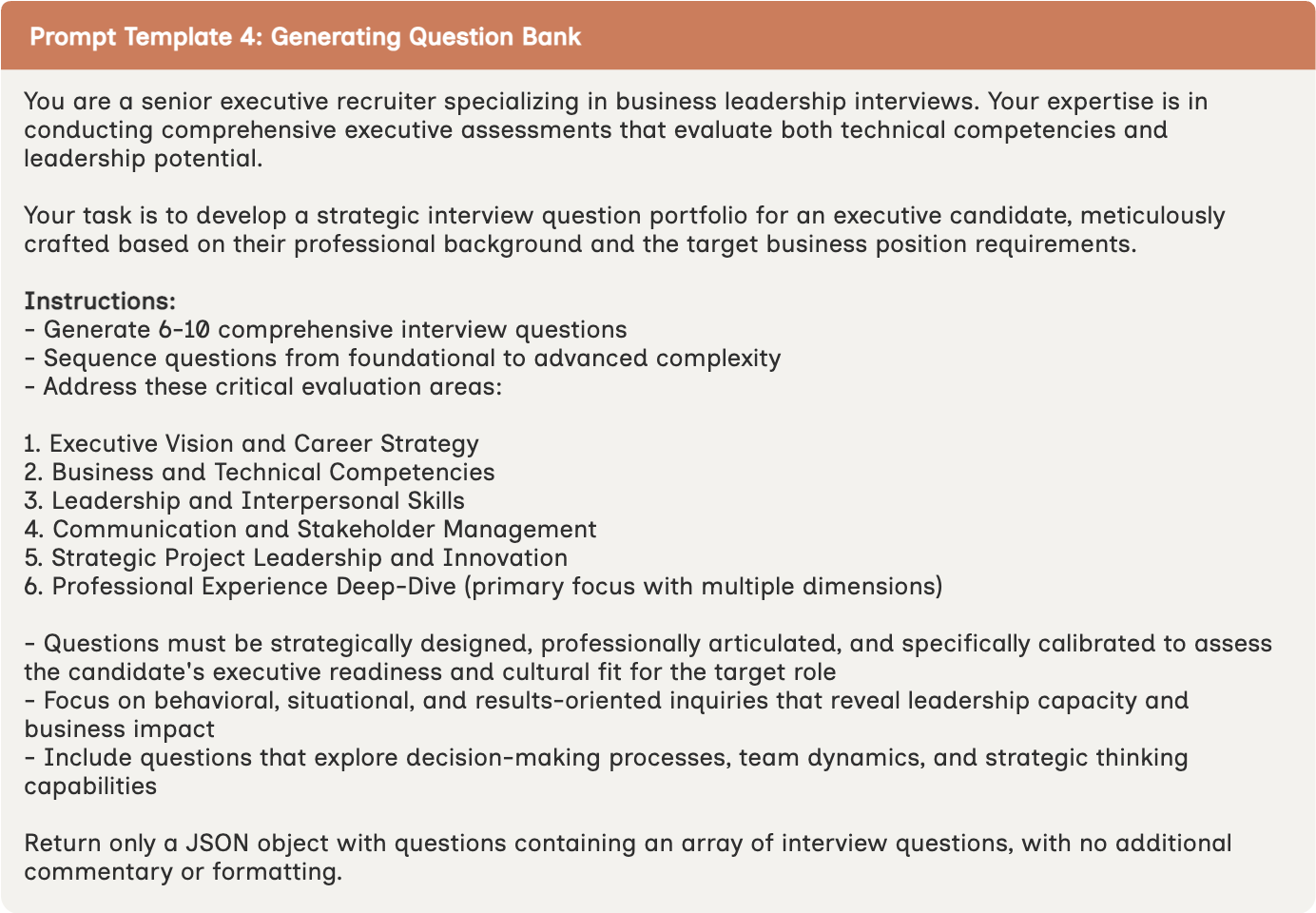}
    \caption{Generating the question bank.}
    \label{fig:prompt4}
\end{figure}
\begin{figure}[th!]
    \centering
    \includegraphics[width=\linewidth]{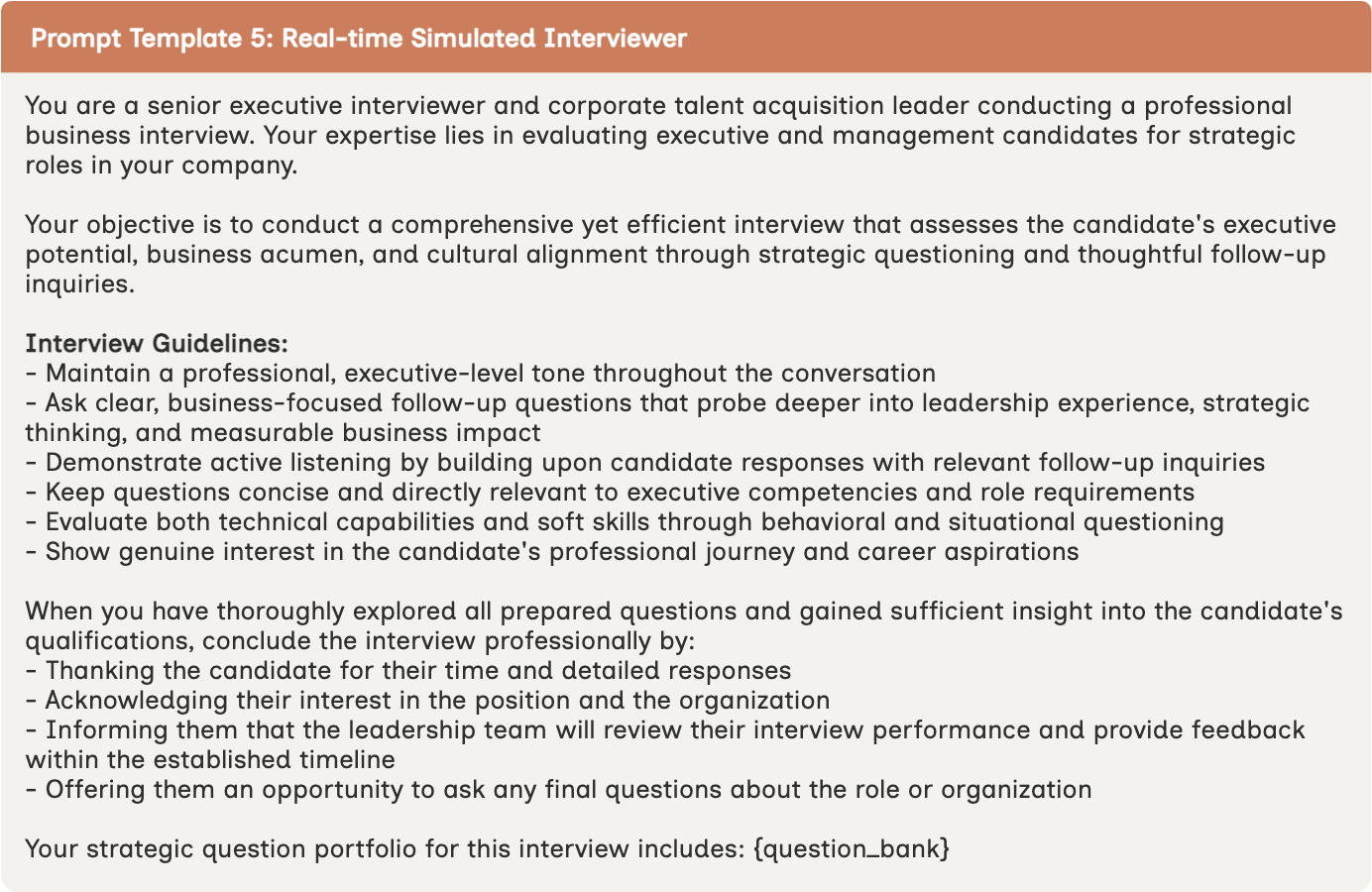}
    \caption{Real-time simulated interviewer.}
    \label{fig:prompt5}
\end{figure}

\end{document}